\documentclass{emulateapj}
\usepackage{upgreek}
\def\spitz{{\it Spitzer }}
\def\wise{WISE~0855-0714}
\def\wist{WISE~1405+5534}

\begin{document}

\shortauthors{Esplin et al.}
\shorttitle{Photometric Monitoring of the Coldest Known Brown Dwarf}

\title{Photometric Monitoring of the Coldest Known Brown Dwarf with the
{\it Spitzer Space Telescope}\altaffilmark{1}}

\author{
T. L. Esplin\altaffilmark{2},
K. L. Luhman\altaffilmark{2,3},
M. C. Cushing\altaffilmark{4},
K. K. Hardegree-Ullman\altaffilmark{4},
J. L. Trucks\altaffilmark{4}.
A. J. Burgasser\altaffilmark{5},
\& A. C. Schneider\altaffilmark{4}}

\altaffiltext{1}{Based on observations made with the {\it Spitzer Space
Telescope}, which is operated by the Jet Propulsion Laboratory,
California Institute of Technology under a contract with NASA.}
\altaffiltext{2}{Department of Astronomy and Astrophysics, The Pennsylvania
State University, University Park, PA 16802; taran.esplin@psu.edu.}
\altaffiltext{3}{Center for Exoplanets and Habitable Worlds,
The Pennsylvania State University, University Park, PA 16802.}
\altaffiltext{4}{The University of Toledo, OH 43606}
\altaffiltext{5}{Center for Astrophysics and Space Science, 
University of California San Diego, La Jolla, CA 92093}

\begin{abstract}
Because WISE J085510.83$-$071442.5 (hereafter \wise) is the coldest known
brown dwarf ($\sim250$~K) and one of the Sun's closest neighbors (2.2~pc),
it offers a unique opportunity for studying a planet-like atmosphere
in an unexplored regime of temperature.
To detect and characterize inhomogeneities in its
atmosphere (e.g., patchy clouds, hot spots), we have performed time-series
photometric monitoring of \wise\ at 3.6 and 4.5~\micron\ with the 
{\it Spitzer Space Telescope} during two 23~hr periods that were
separated by several months.
For both bands, we have detected variability with peak-to-peak amplitudes
of 4--5\% and 3--4\% in the first and second epochs, respectively.
The light curves are semi-periodic in the first epoch for both bands,
but are more irregular in the second epoch.
Models of patchy clouds have predicted a large increase
in mid-IR variability amplitudes (for a given cloud covering fraction)
with the appearance of water ice clouds
at $T_{\rm eff}<$375~K, so if such clouds are responsible for the
variability of \wise, then its small amplitudes of variability indicate
a very small deviation in cloud coverage between hemispheres.
Alternatively, the similarity in mid-IR variability amplitudes between
\wise\ and somewhat warmer T and Y dwarfs may suggest that they share a 
common origin for their variability (i.e., not water clouds).
In addition to our variability data, we have examined other
constraints on the presence of water ice clouds in the atmosphere of \wise,
including the recent mid-IR spectrum from \citet{ske16}. 
We find that robust evidence of such clouds is not yet available.

\end{abstract}

\keywords{brown dwarfs --- infrared: stars ---
solar neighborhood --- stars: low-mass --- planets and satellites: atmospheres}

\section{Introduction}

In multiple temperature regimes for brown dwarfs, condensates are predicted to
form clouds, which can significantly influence the emergent spectra and colors \citep{ack01}. 
The spectra of  L dwarfs \citep[1300--2200~K;][]{ste09} are best fit by models that include a
thick cloud layer of iron, silicates, and corundum \citep{sau08}. 
Those clouds break up non-uniformly and disappear as brown dwarfs grow cooler
and enter the T dwarf sequence  \citep[500--1300~K;][]{ste09}, 
as indicated by the near-infrared (IR) colors \citep{bur02}, 
photometric and spectral variability
\citep{bue14,bur14,rad14,radetal14,wil14,yan16}, 
and surface maps \citep{cro14,kar16} of objects near the L/T transition. 
Clouds may appear again below 900 K based on the colors of late T dwarfs, 
this time in the form of sulfides \citep{mor12}. 
Photometric variability at near-IR wavelengths has been reported in this 
temperature regime, which has been attributed to clouds \citep{yan16}.
Among the Y dwarfs \citep[$<$500~K;][]{dup13}, 
additional clouds of water and ammonia are predicted to form at $<$350~K
and $<$200~K, respectively \citep{burr03,mor14a}.
When water clouds are present, they are expected to be patchy \citep{mor14a}, 
and hence amenable to detection through variability.
The only Y dwarfs with published time-series photometry,
WISE J140518.39+553421.3 (hereafter \wist) and WISEP J173835.52+273258.9,
do exhibit variability but they are likely too warm to have water ice clouds
\citep[$\sim$400~K;][]{cus16,leg16}.

The most promising brown dwarf for the detection of water clouds is  
WISE J085510.83--071442.5 (hereafter \wise).
It is the coldest known brown dwarf \citep[$\sim$250~K;][]{luh14},
making it the most likely one to harbor water clouds. In addition, it is the
fourth closest system to the Sun \citep[2.23$\pm$0.04~pc;][]{luh14,luhes16},
so it is relatively bright for its low luminosity.
As with other Y dwarfs\footnote{\wise\ has not been spectroscopically
classified, but it is very likely to be a Y dwarf based on its luminosity.},
\wise\ is much too faint at near-IR wavelengths for accurate photometric
monitoring \citep{bea14,fah14,kop14,luh14,wri14,luhes16,sch16}.
Currently, such measurements are only feasible in mid-IR bands with the
Infrared Array Camera \citep[IRAC;][]{faz04} on the {\it Spitzer Space
Telescope} \citep{wer04}.

In this paper, we present time-series IRAC photometry of \wise\ during two 23
hour periods. We begin by describing the observations and data reduction
(Section~\ref{sec:obs}). We use these data to characterize the variability of
\wise, which is then compared to the predictions of models that produce
variability through either patchy clouds or hot spots
(Section~\ref{sec:analysis}). We conclude by assessing the evidence of water
ice clouds in the atmosphere of \wise\ from our variability measurements and
previous observations (Section \ref{sec:disc}). 

\section{Observations and Data Reduction}
\label{sec:obs}

In the post-cryogenic mission of {\it Spitzer}, 
IRAC collects data with two operable 256 $\times$ 256 arrays.
Each array has a plate scale of 1$\farcs$2 pixel$^{-1}$ and a field of view of 
$5\farcm2 \times 5\farcm2$.
The arrays simultaneously image adjacent areas of sky in filters
centered at 3.6 and 4.5 $\upmu$m,
which are denoted as [3.6] and [4.5], respectively.
Point sources in the images have FWHM=1$\farcs$7.

To minimize the errors in our time series IRAC photometry of \wise\ due to 
variations in intra-pixel sensitivity \citep{rea05},
the images were taken in the ``staring mode" with the following strategy
\citep{irachpp}: 1) for each of the two filters, the target was placed on
the ``sweet spot" for that array, which is a portion of a pixel near
the corner of the array in which the sensitivity as a function of intra-pixel
position has been well-characterized and
2) prior to collection of the science data, the target was imaged for 30 min
at [4.5] to provide time for the spacecraft pointing to settle.
Following those steps, 
\wise\ was observed continuously during two 23 hour periods   
on 2015 March 10 and 2015 August 3.  During each period, 
we obtained 405 images with exposure times of 96.8 s at [4.5],
which were immediately followed by 405 images with exposures times of 93.6 s at [3.6]. 
These data were collected through Astronomical Observation Requests 52667904,
52668160, 52668672, and 52668928 within program 11056 (K. Luhman).

We began our reduction of the data using the 
Corrected Basic Calibrated Data frames produced by the pipeline
at the \spitz Science Center (SSC).
We measured positions and fluxes of \wise\ in each of those frames with
a point response function (PRF) fitting routine in
the SSC's Astronomical Point source EXtractor \citep[APEX;][]{mak05}, 
which produces more accurate IRAC astrometry than 
other commonly used algorithms \citep{esp16}. 
APEX was used with the default parameters except for a 5$\times$5 pixel
fitting region.  Because aperture photometry has been measured for most
previous time-series data from IRAC, we also applied that method to our
data for comparison to the results of PRF fitting. 
The aperture photometry was measured with {\tt phot} in 
IRAF using an aperture radius of 1.5 pixels and a
background annulus of 3 pixels, which was found to produce the least scatter 
in photometry for \wise\ relative to other annuli.
To compare the data from PRF fitting and aperture photometry in a given band,
we calculated the median absolute deviation (MAD) of the data and
rejected outliers that deviated by $>3\times$MAD.
We then fit the unrejected data using a Nadaraya-Watson regression estimator
as implemented in the function {\tt npregbw} from the {\it np} package 
\citep{hay08} within R \citep{rteam}.
After dividing the data by that fit to remove intrinsic variability,
we recomputed the MAD.
The MADs from APEX and {\tt phot} were similar for [4.5], but APEX produced
significantly lower values at [3.6], where \wise\ is much fainter.

\begin{figure}[h]
	\centering
	\includegraphics[trim = 0mm 0mm 0mm 0mm, clip=true, scale=0.75]{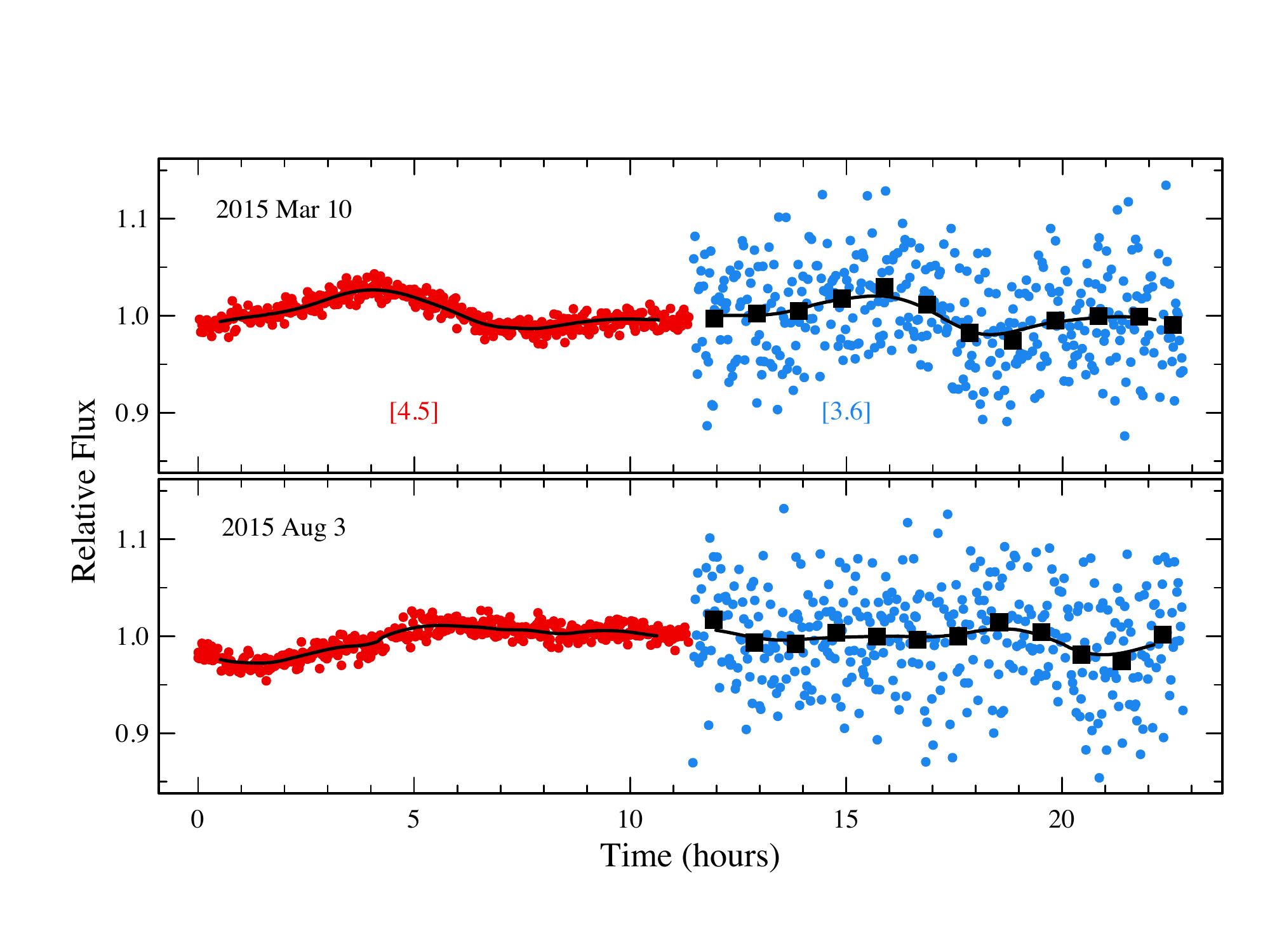}
\caption{
Time-series IRAC photometry of \wise\ during two 23 hour periods.
To illustrate the variability in these data,
we include a {\bf non-parametric} fit to the light curve 
for each band and epoch (black lines) and 
we have binned the [3.6] data into 12 equal time intervals (solid squares).
Measurements that differed from the fits by $>$3$\times$MAD have been omitted.
}
\label{fig:curve}
\end{figure}

We have investigated methods of correcting for systematic noise
in our photometry due to the varying intra-pixel sensitivities.
The SSC has measured high-resolution gain maps of the sweet spots for [3.6] and
[4.5], which can be used for such corrections. 
However, because \wise\ is much fainter at [3.6] than [4.5],
the uncertainties in its positions in individual frames are larger than the 
sweet spot of [3.6] and thus a correction for varying intra-pixel sensitivities
was not possible in  that band.  At [4.5],  the positions of \wise\ drifted
out of the sweet spot for roughly half of the first epoch, but they
remained within it throughout the second epoch. As a result, the latter data 
were suitable for
correction using the SSC's gain map. Applying that correction required that
we reduce those data a second time following the SSC's recommended procedure,
which is performed with the IDL routines {\tt box\_centroider},  {\tt aper},
and {\tt pixel\_phase\_correct\_gauss} \citep{irachpp}.
The application of the gain map predicted a maximum change of $<0.5$\% and
did not produce a lower MAD or any noticeable change to the intrinsic variability compared to the uncorrected photometry from APEX. 
We tested an alternative method of correcting for the varying intra-pixel 
sensitivity from \citet{knu08} and \citet{hei13}, but it also did not reduce
the MADs for either of the two epochs at [4.5] or alter the variability behavior. Therefore, we have adopted the
uncorrected photometry produced by PRF fitting with APEX for both bands and 
epochs. Typical signal to noise ratios (S/N) for the those data 
are 10.2/146.8 and 10.9/142.3 for [3.6]/[4.5] 
in the first and second epochs, respectively.
The APEX data are listed in Tables~\ref{tab:3.6} and \ref{tab:4.5} 
and are plotted as a function of time in Figure \ref{fig:curve}.
We have omitted from Figure \ref{fig:curve} measurements that deviated
by more than 3$\times$MAD from the Nadaraya-Watson regression that
we calculated previously. 

Because IRAC is capable of measuring astrometry with high precision 
\citep{esp16}, it might seem possible to search for perturbations in
the astrometry of \wise\ due to an unseen companion. However, for any
plausible combination of orbital separation and mass ratio, the astrometric
perturbations would be too small to detect during a 12-hour period given the
the typical astrometric errors of 7~mas for \wise\ in individual [4.5] frames
and additional systematic errors among those frames.

\section{Analysis}
\label{sec:analysis}

\subsection{Variability Characteristics}

As shown in Figure \ref{fig:curve},
\wise\ exhibits noticeable variability in both bands and epochs.
For instance, the peak-to-peak amplitudes are 4--5\% at both
[3.6] and [4.5] in the first epoch.
While the light curve of the first epoch appears semi-periodic in both bands,
the shape does not follow a single-period sinusoid, 
which suggests that the mechanism producing the 
variability evolves on the timescale of hours
or is spatially complex (i.e., several spots). 
The second epoch also shows variability in both filters but is less sinusoidal
and has a lower amplitude (3--4\%).  The differences between the
two epochs indicate that the light curve evolves on a timescale of months.
These IRAC data have provided the third detection of variability in a
(likely) Y dwarf \citep{cus16,leg16}. The light curves for \wise\ are
roughly similar to those of the previously studied Y dwarfs
in terms of amplitudes, but they are less periodic.
In addition to our two epochs of time-series photometry,
\wise\ has been briefly imaged with {\it Spitzer} on several other occasions 
across a period of two years \citep{luh14,mel15,luhes16}.
Among all of the available photometry, the [3.6] and [4.5] data have
spanned ranges of $\sim$0.16 and 0.13~mag, respectively.

Our attempts to identify a single rotational period for \wise\ have
produced inconclusive results.
The peaks in power in Lomb-Scargle periodograms \citep{lom76,sca82}
appear at 6.8/9.3 and 9.0/5.3 hrs for [3.6]/[4.5] at the first and 
second epochs, respectively. 
We also fit the data in each band and epoch to a double sine model
where the second sine has a period twice that of the first \citep{cus16}.
While this model produces random residuals, indicating a good fit, 
the predicted periods are only consistent between epochs at the two sigma level.
Specifically, they have values of $9.7^{+0.9}_{-0.8}$/$10.8^{+0.7}_{-0.7}$ and 
$14^{+2}_{-2}$/$13.3^{+0.5}_{-0.4}$~hrs for [3.6]/[4.5] 
in the first and second epochs, respectively.
\cite{man15} successfully estimated a rotation period for 
a brown dwarf with a rapidly changing light curve with a more complicated model,
but that model fails to converge on a solution for \wise. 
Additional time-series data would be needed to reliably measure the rotation 
period and the relative phase between the light curves of the two bands. 

Because the spectra of cold brown dwarfs are non-Planckian,
different wavelengths can sample different pressure levels in the atmosphere.
In the case of \wise, the similarity between the [3.6] and [4.5] light
curves at both epochs indicate that they may be sampling similar levels.
However, without a measurement of a relative phase between the light curves of
the two bands, it is difficult to make strong predictions about the 
structure of the atmosphere of \wise. For example, 
an absence of a phase offset between bands would indicate 
that dynamical circulation is efficiently carrying thermal energy 
through these pressure levels.

Photometric variability in brown dwarfs could potentially arise from
a number of spot-producing mechanisms, including magnetic activity, 
atmospheric chemical abundance variations, non-uniform cloud coverage,
and variable temperature profiles causing hot/cold spots.
In the following two sections, we discuss whether the variability of
\wise\ can be explained by the latter two mechanisms.
We have not considered the first two mechanisms because 
light curves have not been modeled for an atmosphere with 
heterogenous chemical abundances and because
magnetically-induced starspots probably do not form
in the neutral atmospheres of the coldest brown dwarfs,
even in the presence of magnetic fields \citep[e.g.,][]{moh02}.

\subsection{Patchy Water Clouds}
\label{sec:cloud}

\begin{figure}[h]
	\centering
	\includegraphics[trim = 0mm 0mm 0mm 0mm, clip=true, scale=1]{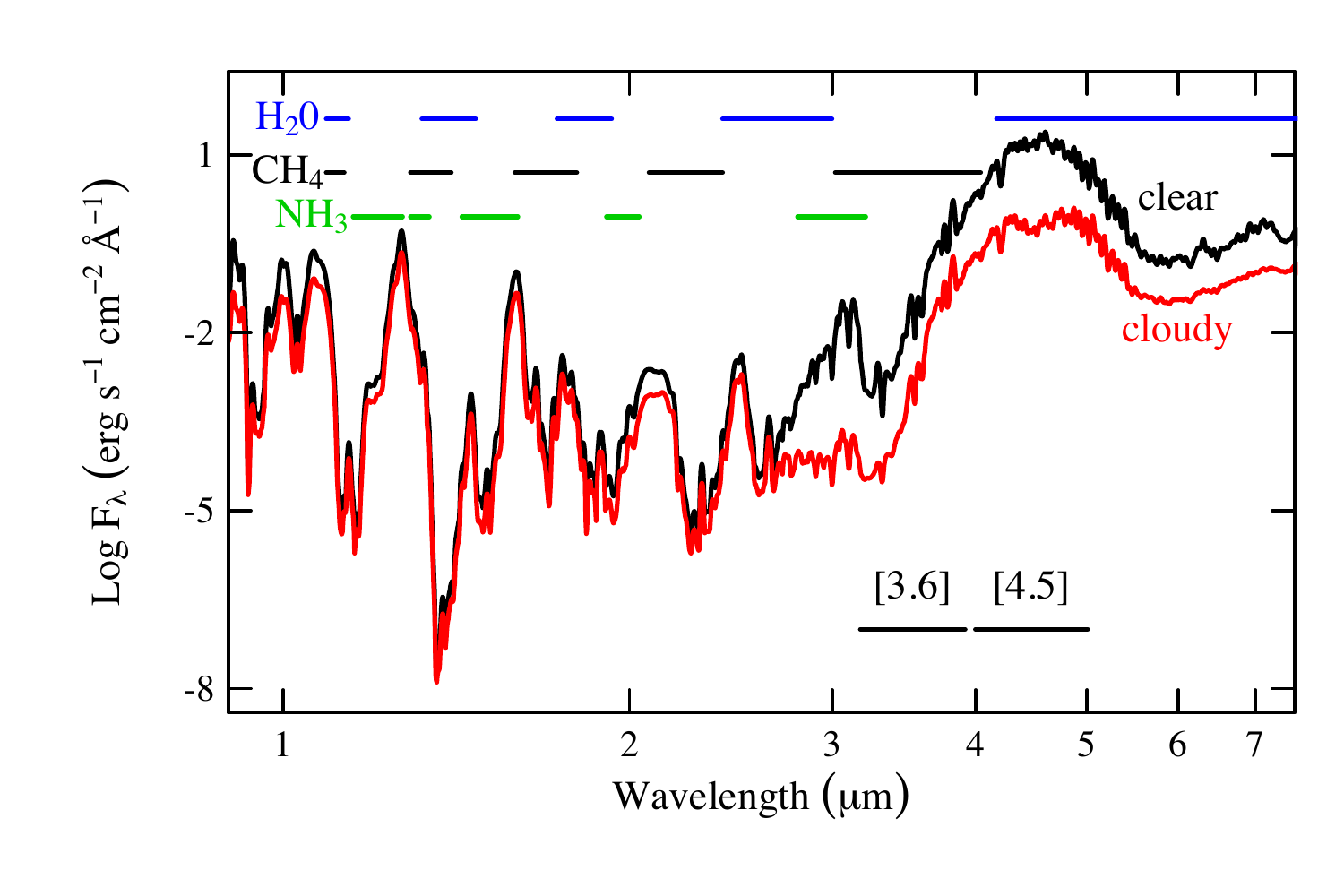}
\caption{
Model spectra for clear (black) and cloudy (red) columns in the atmosphere of a  
brown dwarf with $T_{\rm eff}$=250 K, log $g$=4.0 cm s$^{-2}$, $f_{\rm sed}$=3,
and a cloud coverage of 50\% \citep[$h=0.5$;][]{mor14a}.
The wavelengths of prominent molecular absorption bands and the [3.6] and [4.5]
filters are indicated.
Longward of $\sim$3~$\upmu$m, 
water clouds are predicted to significantly reduce the emergent flux.
Any deviation of the cloud coverage fraction between hemispheres 
would produce rotationally-modulated variability at [3.6] and [4.5].
}
\label{fig:spec}
\end{figure}

If water clouds are present in the atmospheres of brown dwarfs as cold as
\wise, those clouds are likely to be patchy rather than uniform \citep{mor14a}.
To construct a self-consistent and stable model with non-uniform cloud coverage,
\cite{mor14a} computed emergent spectra for cloudy and clear atmospheric 
columns with a single temperature-pressure profile and then combined the
fluxes ($F_\nu$) from the two columns in the following manner: \begin{equation}
\label{equ:cloudy}
F_{\rm \nu,total} = hF_{\rm \nu,clear} + (1-h)F_{\rm \nu,cloudy},
\end{equation}
where $h$ is the fraction of the atmosphere without clouds \citep{mar10}.
Any deviation of $h$ between hemispheres from the global average 
of a model for \wise\ would produce photometric variability. In addition,
these patchy cloud models predict substantially different emergent flux
between the clear and cloudy columns at the two IRAC bands. 
For example, in Figure \ref{fig:spec} we show the spectra of both columns for 
a model brown dwarf with $T_{\rm eff}$=250 K, log $g$=4.0 cm s$^{-2}$,
$f_{\rm sed}\footnote{
This parameter describes the efficiency of cloud particle growth, or
sedimentation \citep{ack01}, where higher values correspond to larger particle
sizes and consequently geometrically thinner clouds.}$=3 and a cloud coverage 
of 50\% ($h$= 0.5) \citep{mor14a}\footnote{ 
The combined spectra from the cloudy and clear columns in the models from
\cite{mor14a} are available at
{\url www.ucolick.org/$\sim$cmorley/cmorley/Models.html}.
We have made use of the separate spectra from those columns, which were
provided by C. Morley.}.
The cloudy column exhibits much lower flux longward of $\sim$3~$\upmu$m while
the near-IR spectrum is affected less by the presence of water clouds.
As a result, our IRAC observations are ideal for detecting photometric variability produced from water clouds.

\begin{figure}[h]
	\centering
	\includegraphics[trim = 0mm 0mm 0mm 0mm, clip=true, scale=1]{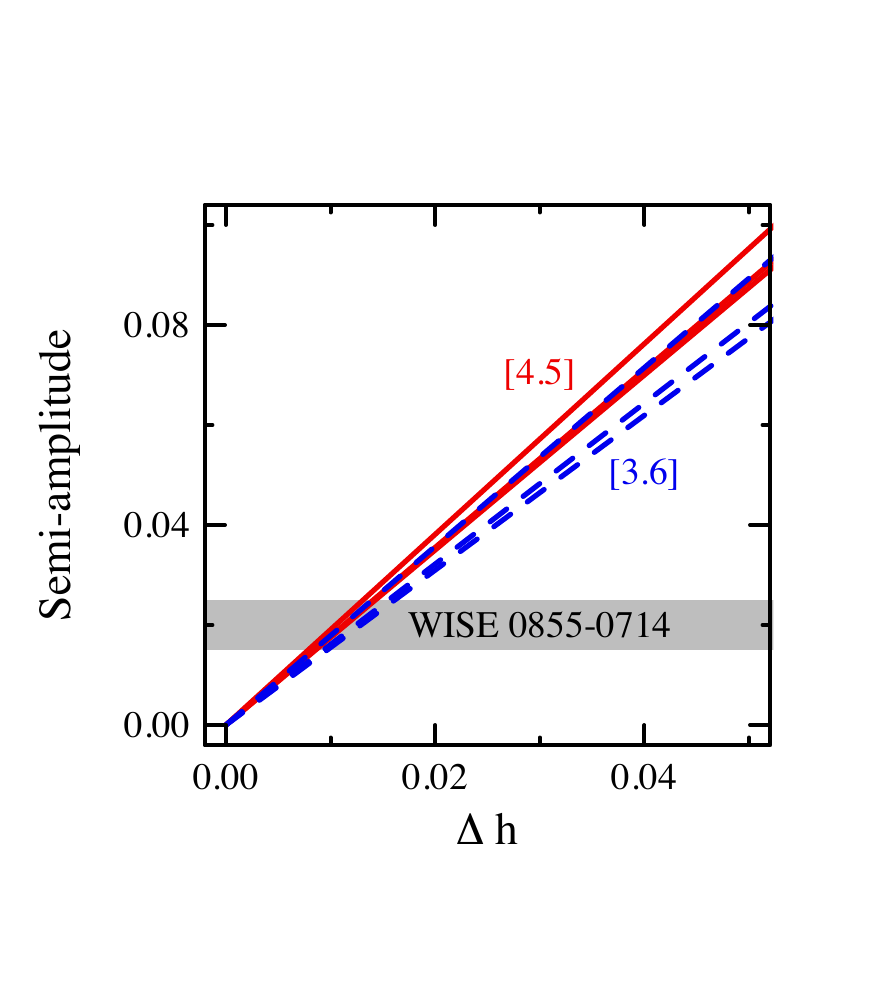}
\caption{
Predicted semi-amplitudes of variability for [3.6] (blue) and [4.5] (red) 
assuming a deviation in the cloud coverage fraction of $\Delta h$
from a global average ($h$=0.5) between hemispheres for the model
from Figure~\ref{fig:spec} \citep{mor14a} {\bf with $f_{\rm sed}$=3, 5, and 7,
which correspond to the middle, top, and bottom lines for each band,
respectively.} The observed 
semi-amplitudes for the two bands are similar for \wise\ (gray region).
Only a 1\% deviation in the cloud coverage is needed to explain
the observations. 
}
\label{fig:cloud}
\end{figure}

To test whether water clouds can produce the observed variability in 
\wise, we follow the prescription of \cite{cus06,cus16}.
Using the $h$=0.5 model of Figure \ref{fig:spec},
we assumed that the cloud cover between hemispheres deviates from the global
average by $\Delta h$ and calculated predicted fluxes for
$F_{\nu,h+\Delta h}$ and $F_{\nu,h-\Delta h}$ from Equation \ref{equ:cloudy}.
This deviation would produce variability with a semi-amplitude of 
\begin{equation}
A_\lambda = \frac{\rm{max} \left[ F_{\nu,h+\Delta h} (\lambda),F_{\nu,h-\Delta h} (\lambda) \right] }{\rm{average} \left[ F_{\nu,h+\Delta h} (\lambda),F_{\nu,h-\Delta h} (\lambda) \right] } - 1.
\end{equation}
In Figure \ref{fig:cloud}, we show the predicted semi-amplitudes at [3.6]
and [4.5] as a function of $\Delta h$ for 
$T_{\rm eff}$=250 K, log $g$=4.0 cm s$^{-2}$, cloud coverage 
of 50\%, and $f_{\rm sed}$=3, 5 and 7.
Amplitudes in the two bands do not differ significantly for $\Delta h < 0.05$ 
and increase linearly with increasing deviation from homogeneous cloud cover.
We find that only a small deviation from a global average $h$ 
($\Delta h \approx 0.01$) is needed to reproduce the observed amplitudes
of variation in both bands for \wise.  
Although this model of patchy clouds is able to reproduce the observed
variability of \wise, it may not be the best explanation of those data,
as discussed in Section \ref{sec:disc}.

\subsection{Hot Spots}
\label{sec:hotspot}

A brown dwarf's photometric variability could be produced by 
temperature perturbations in its atmosphere, which would be 
manifested as time-varying hot and/or cold spots in the photosphere. 
Such perturbations might occur if the atmosphere circulates faster
than the gas can equilibrate or if the photosphere is radiatively coupled
to changes at deeper pressure levels,
such as deep heterogeneous clouds \citep{sho13,rob14}.
\cite{mor14b} simulated the effect of the latter mechanism by
injecting energy at various pressure levels into cloudless static models
for brown dwarfs with temperatures of 400--1000 K. \citet{cus16}
performed the same exercise for 500~K in an attempt to reproduce
the variability of \wist.
The two studies found that spots produce large variability in strong absorption
features like the CH$_4$ band within the [3.6] band, whereas the variability
was smaller at wavelengths with less absorption, as in the case of [4.5].
In contrast, \wise\ exhibits similar variability amplitudes in those two bands.
Thus, it appears unlikely that hot spots are the cause of its variability.

\section{Discussion}
\label{sec:disc}

Previous studies have attempted to constrain the presence of water ice
clouds in the atmosphere of \wise\ using photometry and spectroscopy.
\citet{fah14} reported a possible 2.6~$\sigma$ detection of \wise\ in
a medium-band filter within the $J$ band.
Those data were used to place the object in a diagram of $M_{W2}$ versus
$J-W2$, where $W2$ is a band from the {\it Wide-field Infrared Survey
Explorer} \citep{wri10} that is similar to [4.5] from {\it Spitzer}.
The position of \wise\ in that diagram was better reproduced by
cloudy models than cloudless models \citep{mor12,mor14a,sau12}, which
was interpreted as evidence of water ice clouds.
However, \citet{luhes14} demonstrated that \wise\ was roughly midway between
those cloudless and cloudy models in a similar diagram of $M_{4.5}$ versus
$J-[4.5]$, and that its position was best matched by cloudless models that
employed non-equilibrium chemistry \citep{sau08,sau12}.
After measuring photometry for \wise\ in several additional near-IR bands,
\cite{sch16} and \cite{luhes16} found that no single suite of models
provided a clearly superior match to the observed spectral energy distribution
(SED), and that all of the models differed significantly from the data.
Thus, the photometry and models that are currently available do not
provide any indication of whether water ice clouds are present in \wise.

A spectroscopic investigation of water ice clouds in \wise\ has been recently
performed by \citet{ske16}. They obtained the only spectrum to date of
the brown dwarf, which spans from 4.5--5.1~\micron. Although its resolution
and S/N were low, the spectrum exhibited absorption features that appeared
to be statistically significant, many of which coincided with features in
model spectra computed by \citet{ske16}. 
The spectra predicted by cloudless and partly cloudy models
were indistinguishable for the wavelength range and resolution of the data.
The atmospheric temperature structure for a brown dwarf near the temperature
of \wise\ does not converge with full coverage of water clouds \citep{mor14a},
so \citet{ske16} developed a simplified model for that scenario, which used
a gray, fully absorbing cloud with no specified composition. The cloud-top
pressure in that model was varied to optimize the match to the observed
spectrum. The resulting best-fit spectrum agreed somewhat better with
the data than the spectra from the cloudless and partly cloudy models, which
was cited as a detection of clouds. Given the temperature of \wise, it
is expected that such clouds would be composed of water \citep{burr03,mor14a}.

It is unclear whether the spectrum from \citet{ske16} actually contains
evidence of water clouds.
The fully cloudy model from that study considered gray absorbers, whereas
water ice is non-gray across the wavelength range of the spectrum of 
\wise\ \citep{mor14a}, and it assumed that the cloud coverage is uniform,
which is not expected for water clouds \citep{mor14a}.
In addition, models with patchy clouds have self-consistent temperature
structures, which has not been possible for the fully cloudy models
\citep{mor14a}.
Given the variety of uncertainties in models of the coldest
brown dwarfs \citep{mor14a} and the large differences between the observed
and predicted SEDs for \wise\ \citep{sch16,luhes16},
the current models may not be sufficiently accurate for subtle
differences in predicted spectra to provide meaningful insight into the
physical properties of \wise.

Our variability measurements can provide additional constraints on
the existence of water ice clouds in the atmosphere of \wise.
According to the partly cloudy models of \citet{mor14b},
cloud-induced variability at mid-IR wavelengths should become much larger
(for a given cloud covering fraction)
at temperatures below $\sim375$~K with the onset of water clouds.
The amplitudes that we have measured for \wise\ ($\sim250$~K) are only
a few percent, which would require very small deviations in cloud coverage
between hemispheres ($\Delta h\sim0.01$) if water clouds are responsible
for the variability.
Meanwhile, the mid-IR amplitudes for \wise\ are
similar to those observed for early Y dwarfs
\citep[$\sim400$~K,][]{cus16,leg16} and late T dwarfs
\citep[800--1000~K,][]{met15,yan16}, which should be too warm to harbor
water clouds. If water clouds are producing the variability of \wise,
then it must coincidentally have
a value of $\Delta h$ that produces roughly the same amplitudes as
the different variability mechanism that operates in those objects.
Alternatively, the similarity in the amplitudes
of \wise\ and somewhat warmer brown dwarfs may indicate that they share
a common origin for their variability (i.e., not water clouds).
Based on these results and the previous work that we have discussed,
we conclude that robust evidence of water clouds in the atmosphere
of \wise\ is not yet available.

\acknowledgements

We thank Caroline Morley and Mark Marley for providing their model calculations.
We acknowledge support from a grant from NASA issued by the Jet Propulsion
Laboratory (JPL), California Institute of Technology. The {\it Spitzer
Space Telescope} is operated by JPL/Caltech under a contract with NASA.
The Center for Exoplanets and Habitable Worlds is supported by the
Pennsylvania State University, the Eberly College of Science, and the
Pennsylvania Space Grant Consortium.

\clearpage

\begin{deluxetable}{cccc}
\tabletypesize{\scriptsize}
\tablewidth{0pt}
\tablecaption{Time-series Photometry for \wise\ in [3.6]\label{tab:3.6}}
\tablehead{
\colhead{HMJD} &
\colhead{[3.6]} &
\colhead{error} & 
\colhead{Outlier?} \\
\colhead{} & \colhead{(mag)} & \colhead{(mag)} & \colhead{}}
\startdata
57091.0253 & 17.669 &  0.099 & Y \\
57091.0264 & 17.282 & 0.066 & N \\
57091.0276 & 17.258 & 0.064 & N \\
57091.0288 & 17.380 & 0.071 & N \\
57091.0300 & 17.411 & 0.073 & N 
  \enddata
\tablecomments{
Photometry is absent for four frames because APEX failed to converge
on a measurement. This table is available in its entirety in a machine-readable
format.  A portion is shown here for guidance regarding its form and content.}
\end{deluxetable}

\begin{deluxetable}{cccc}
\tabletypesize{\scriptsize}
\tablewidth{0pt}
\tablecaption{Time-series Photometry for \wise\ in [4.5]\label{tab:4.5}}
\tablehead{
\colhead{HMJD} &
\colhead{[4.5]} &
\colhead{error} & 
\colhead{Outlier?} \\
\colhead{} & \colhead{(mag)} & \colhead{(mag)} & \colhead{}}
\startdata
57090.5488 & 13.876 & 0.008 & Y \\
57090.5500 & 13.840 & 0.007 & N \\
57090.5512 & 13.854 & 0.007 & N \\
57090.5523 & 13.846 & 0.007 & N \\
57090.5535 & 13.855 & 0.007 & N
\enddata
\tablecomments{
Photometry is absent for one frame because APEX failed to converge
on a measurement. This table is available in its entirety in a machine-readable
format.  A portion is shown here for guidance regarding its form and content.}
\end{deluxetable}

\end{document}